\documentclass{aa}
\usepackage{graphicx}
\begin{document}

   \title{Illuminating the darkness}

   \subtitle{More evidence from faint-star proper-motions for a cool
and warm component to the local dark matter in the Galaxy
\thanks{Based on observations collected with the Hubble Space
Telescope, which is operated by AURA, Inc., under contract with the
National Science Foundation.}}

   \author{Ren\'e A. M\'endez}

   \offprints{Ren\'e A. M\'endez}

   \institute{European Southern Observatory, Casilla 19001, Santiago
   19, Chile \email{rmendez@eso.org}}

   \date{Received April 24, 2002; accepted July 16, 2002}

   \abstract{We present new evidence, based on faint {\it HST}
proper-motions, for a bi-modal kinematic population of old white
dwarfs, representative of the Thick-Disk and Halo of our Galaxy. This
evidence supports the idea of a massive Halo comprised of faint and
old white dwarfs, along with an extant population of Thick-Disk white
dwarfs. We show how most of the required dark matter in the solar
vicinity can be accounted for by the remnants from these two
components together.

    }

   \maketitle


\section{Introduction}

The claim by Oppenheimer et al. (\cite{oppenheimer}) that they had
found a significant population of Halo stars from a kinematic survey
toward the South Galactic Pole opened up an interesting discussion
regarding the nature of dark matter in the solar neighborhood. Their
discovery seemed to corroborate earlier findings using very deep {\it
HST} photometry that also pointed out to the existence of an as yet
unobserved component to the Halo, identified as very cool and ancient
white dwarfs (M\'endez and Minniti 2000). However, several authors
have indicated that Oppenheimer's sample could be also interpreted as
the tail of a 'warmer' white dwarf component (in the sense of having
shorter cooling ages and therefore higher surface temperatures),
better ascribed to the intermediate Thick-Disk population of the
Galaxy (see e.g., Reid et al. 2001; Reyl\'e et al. 2001).

Obviously, disentangling the true nature of Oppenheimer's objects is
important to understand their contribution to the dark matter problem
in the Galaxy. In this paper we use {\it HST} data, deep photometry
and proper-motions, to demonstrate that there is indeed a very
important population of what appears to be Thick-Disk white dwarfs
with a large mass density in the solar vicinity. These Thick-Disk
objects could not have been identified before on the Hubble Deep
Fields South and North due to their high Galactic latitude ($|b| \ge
50^o$), which prevented the appearance on these fields-of-view of an
important population of these objects, characterized by a steep
concentration toward the Galactic plane.

A fundamental discovery presented in this paper is that, regardless of
whether the objects found are Thick-Disk or Halo white dawrfs, they
together account for most of the dark matter in the solar
neighborhood.


\section{Description of the photometric and kinematic sample}

Our data is based on a proper-motion membership study of the nearest
Globular Cluster NGC~6397 by King et al.\ (\cite{king}, KACP98
thereafter). They used {\it HST} proper-motions to segregate cluster
members from field stars in order to define, characterize, and study
the faint end of the main sequence on the cluster, in a seminal paper
using {\it HST's} WFPC accurate relative astrometry. Since their main
science goal was to study the cluster, the field stars received
comparatively little attention in that paper. In a previous paper they
had also used the first-epoch exposures to study the cluster's white
dwarf sequence (Cool et al. 1996).

More recently, Castellani et al. (\cite{castellani}) used King's data
set to compare the star- and color-counts with Galactic models,
constraining a number of large-scale Galactic parameters, but no
attention was given to the kinematic information, as derived from the
proper-motions.

In this paper,we use the combined photometric and astrometric
information to characterize, specifically, the blue-faint objects
found in the field population of the KACP98 sample. The main
motivation is the recent discovery of spectroscopically-confirmed old
white dwarfs indicative of a rather massive Galactic component
comprised of these remnants, as previously suggested by the
photometric study of the stellar content on the Hubble Deep Fields
South and North by M\'endez et al.\ (\cite{mendez96b}), and M\'endez
and Minniti (\cite{mendez00a}, MM00 henceforth).

Relative proper-motions for all field and cluster stars in a field
close to the cluster NGC~6397 were derived by KACP98 using an earlier
version of the methods described in Anderson and King (2000, and
references therein). The final proper-motions for field stars are
motions relative to a reference frame in which the cluster is at
rest. Since the internal velocity dispersion for the cluster ($\le
5$~km~s$^{-1}$ or 0.4~mas~y$^{-1}$, where mas stands for milli-arcsec)
is smaller than the proper-motions detected, at the distance of the
cluster, the width of the cluster proper-motion provides a good
indication of the uncertainties in the derived relative motions.

The full sample was segregated by KACP98 into three sub-samples, based
on a proper-motion and photometric selection criteria: NGC~6397
main-sequence stars (1,386 stars), non-main sequence members (64
stars), and generic field stars (929 stars) on a 4.5 arcmin$^2$
field-of-view located at $(l,b)=(338.1^\circ, -12.0^\circ)$,
encompassing the range $15 \le I \le 25$ and $0 \le V-I \le
4.0$. Here, we will concentrate our attention on the field stars of
the KACP98 sample.

The uniqueness of this data set is the availability of proper-motions
for a a sample of very faint stars in a field with modest Galactic
latitude, which then samples a significant path through the Thick-Disk
of the Galaxy (M\'endez \& van Altena 1996). Another fortunate
coincidence is that, unlike other studies being carried out with {\it
HST} on different Globular clusters, which focus on the inner core of
the clusters, this field was acquired toward the outer envelope of
NGC~6397, some 4.5~arcmin from the cluster center, far from the
central 20~arcsec dominated by the cluster's power-law cusp, and
therefore minimizing the 'contamination' from cluster stars.

\subsection{Photometric calibration}

KACP98 published photometry in the F555W and F814W bands of the
'flight-system', as defined by Holtzman et al. (\cite{holtzman},
Ho95). In order to compare our photometry with models, we need to
convert this photometry to a standard system. However, because we only
have magnitudes in two bands we can not apply the iterative method
used by M\'endez et al (\cite{mendez96b}), or MM00, to derive
calibrated photometry. Instead, in this paper, the calibration is done
on a two-step process, by adjusting the zero-points of the
flight-system (equation~\ref{eqn1} below), then by applying the
flight-to-ground transformations (equation~\ref{eqn2} which yields
equation~\ref{eqn3} below), and then explicitly solving for a
quadratic equation on the standard photometry.

Specifically, the mags given by KACP98 are defined by (equation~(7) of
Ho95):

\begin{equation}
WFPC2 = -2.5 \log (DN \, s^{-1}) + Z_{FG}  + 2.5 \log GR_i
\label{eqn1}
\end{equation}
where WFPC2 is the flight magnitude, DN is the measured flux per unit
time on the camera, $Z_{FG}$ are the zero-points as given in Table~6
of Ho95, and $GR_i$ is the detector's gain (slightly different for all
three chips, hence the subscript, in our case this is $\sim$7 $e^-$
ADU$^{-1}$).

On the other hand, the flight-to-ground transformation is accomplished
by equation~(8) on Ho95, which we repeat here for completeness:

\begin{eqnarray}
SMAG & = & -2.5 \log (DN \, s^{-1}) + Z_{FS} + 2.5 \log GR_i + \nonumber \\ 
& & T_{1,FS} \times SCOL + T_{2,FS} \times SCOL^2
\label{eqn2}
\end{eqnarray}
where SMAG and SCOL are the standard-system magnitudes and colors
respectively, $T_1$ and $T_2$ are linear and quadratic coefficients
determined empirically which, along with the zero-points, are given in
Table~7 of Ho95. Combining equations~\ref{eqn1} and~\ref{eqn2} above
thus yields:

\begin{eqnarray}
SMAG & = & WFPC2 + (Z_{FS} - Z_{FG}) + 2.5 \log GR_i + \nonumber \\
& & T_{1,FS} \times SCOL + T_{2,FS} \times SCOL^2
\label{eqn3}
\end{eqnarray}
In our case we have magnitudes in two passbands only, so that SMAG
could be either V or I (the closest match to the F555W and F814W
passbands, see Ho95). Using equation~\ref{eqn3}, and subtracting the
terms for the I band from those of the V band, we end up with:

\begin{eqnarray}
V-I \cor SCOL & = & (F555W - F814W) +  \nonumber \\
& & (T_{1,FS,F5Q5W} - T_{1,FS,F814W}) \times SCOL +  \nonumber \\
& & (T_{2,FS,F555W} - T_{2,FS,F814W})  \times SCOL^2  \nonumber
\end{eqnarray}

which is a quadratic equation on SCOL. Solving for SCOL yields thus
the standard color, which, combined with equation~\ref{eqn3} yields
also the individual apparent magnitudes. There is no iteration
necessary, and the standard magnitudes are uniquely determined (this
is not the case if we have more than two bands, in this case the best
solution is obtained by the iterative process, which best satisfies
(in a minimum-residual sense) equation~\ref{eqn2} simultaneously, for
all passbands). The corrections from the instrumental to the standard
system are, in any case, quite small. For example for F14W=~15.87, and
F555W-F814W=~1.03, one finds that I=~15.83 and V-I=~1.043. Because
these corrections are quite small, errors in the transformation (which
can be up to 0.1~mag, Ho95) are minimal (i.e., we could have as well
used the instrumental magnitudes). The uncertainty in the final
calibrated magnitudes is estimated to be less than 0.05~mag (Cool et
al. 1996).

\subsection{Proper-motions}\label{pms}

The proper-motions given by KACP98 are actually displacements on a
32-month baseline with respect to the mean displacement of the
bona-fide cluster stars. These displacements were converted to actual
relative proper-motions using the WFC plate-scale, and the
baseline. This proper-motions were then converted to motions in RA and
DEC by applying the position angle of the observations. As mentioned
before, the motions are measured with respect to a system that has a
zero mean motion for the cluster. In order to compute true absolute
motions we need to correct for the proper-motion of the cluster, i.e.:

\[
\vec \mu^f_{rel} = \vec \mu^f_{abs} - \vec \mu^{cl}_{abs}
\]
where $\vec \mu^f_{rel}$ are the observed relative motions for field
stars with respect to the cluster, $\vec \mu^{cl}_{abs}$ is the
absolute motion of the cluster, and $\vec \mu^f_{abs}$ is the absolute
motion of the field stars. Fortunately, the motion for NGC~6397 is
known, and it is given in the comprehensive compilation by Dinescu et
al. (\cite{dinescu}), which has $\mu_{\alpha_{abs}}^{cl} \cos \delta =
3.30 \pm 0.50$~mas~y$^{-1}$, and $\mu_{\delta_{abs}}^{cl} = -15.20 \pm
0.60$~mas~y$^{-1}$. We finally converted the motion to the Galactic
system for an easier interpretation of the results. In this system,
the mean absolute motion for the cluster would be $\mu_{l_{abs}}^{cl}
= -11.46$~mas~y$^{-1}$, and $\mu_{b_{abs}}^{cl} =
-10.52$~mas~y$^{-1}$.


\section{Analysis of the sample}

\begin{figure*}
\centering
\includegraphics[width=17cm]{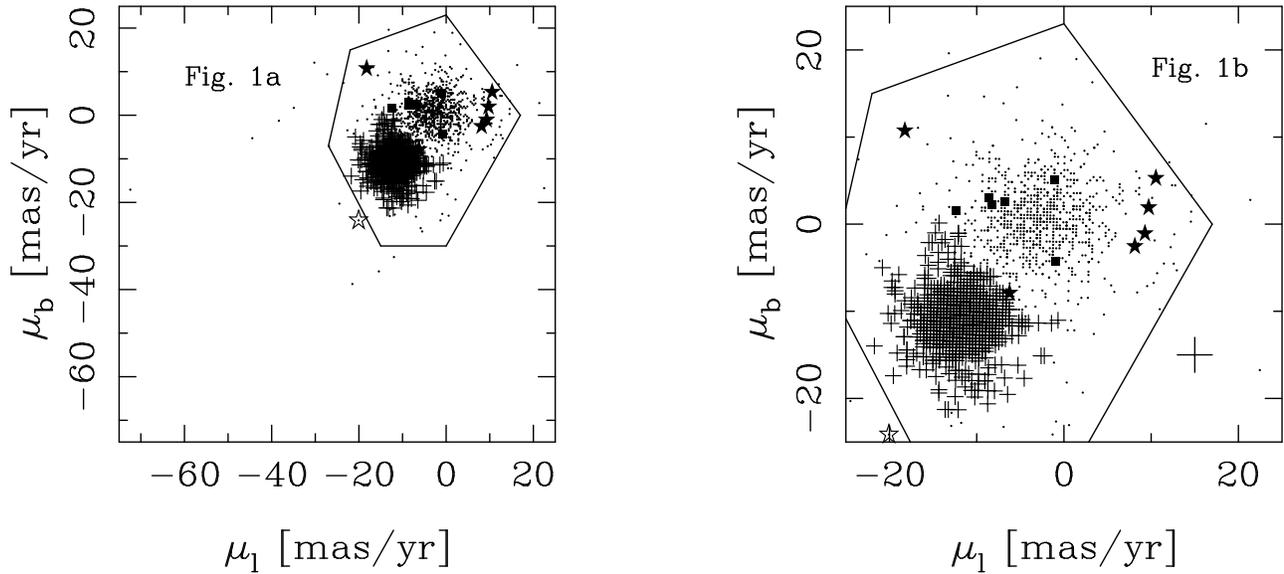}
\caption{Vector point diagram for the whole sample (a), and for an
enlargement (b), showing the most interesting objects found in the
field population. The outliers boundary (solid-line polygon) has been
drawn by eye. The objects outside the polygon are marked by an open
circle in Figs.~\ref{rpmfield} and~\ref{cmd}. Pluses are for cluster
stars, dots for field stars. Among these we have the extremely red
objects (solid squares), and the blue faint population (filled-star
symbol, six stars, and open star symbol, one star). Typical
proper-motions uncertainties ($\sim$~2~mas~y$^{-1}$) are marked by the
large plus symbol in Fig.~1b.}
\label{vpd}
\end{figure*}

The vector point diagram (VPD) for the whole sample is shown in
Fig.~\ref{vpd}. Fig.~1a includes all the data, whereas Fig.~1b shows a
detail of the inner core of the proper-motion distribution. The pluses
indicate the cluster members, whereas the small black dots indicate
the non-members. In addition, several other interesting objects have
been marked by the solid and open star symbols, as well as the filled
square symbol. These objects will be discussed further below. This
figure clearly shows that there is a good separation between cluster
and field stars, although the boundary, specially in the interface
between the two distributions, is somewhat blurry (fortunately this is
of no consequence to our analysis).

Table~\ref{proppm} indicates the main properties of the proper-motion
distributions shown in Fig.~\ref{vpd}, split into Main-sequence
cluster members, and non-members. The method to compute these
parameters, accounting for outliers, has been described in M\'endez et
al. (\cite{mendez00b}).

\begin{table}
\caption{Properties of the bulk proper-motions in Galactic longitude
and latitude for NGC~6397 main-sequence cluster members and field
stars.}
\label{proppm}
\begin{tabular}{@{}ccc}
\hline
\hline
Parameter  & MS cluster members & Field stars  \\
Gal. Longitude  &  mas y$^{-1}$      & mas y$^{-1}$ \\
\hline
Mean       &  -11.86            & -2.88        \\
Mean error &    0.06            &  0.14        \\
   & & \\
Dispersion &    2.06            &  4.30        \\
Disp error &    0.06            &  0.22        \\
\hline
Gal. Latitude   &                    &              \\
Mean       &  -10.78            &  0.62       \\
Mean error &    0.05            &  0.14        \\
   & & \\
Dispersion &    1.91            &  4.04        \\
Disp error &    0.05            &  0.20        \\
\hline
\hline
\end{tabular}
\end{table}

From Table~\ref{proppm} we can estimate the accuracy of the
proper-motions directly from the dispersion in the cluster members
(which should actually have a very small dispersion if there were no
observational errors). We obtain an overall uncertainty of about
2.0~mas~y$^{-1}$ in both coordinates. We also note that the mean
motion for the cluster is what we would have expected from the
absolute motion of the cluster (see Section~\ref{pms}), after
correction from relative to absolute, and within the observational
errors as given on the second row of the table above. While the above
values are for the entire sample, there is a slight increase in
proper-motion errors, as judged from the width of the cluster
distribution, when going to fainter magnitudes, for example while the
dispersion is 1.99 and 1.88~mas y$^{-1}$ in $\mu_l$ and $\mu_b$
respectively for $16 \le I < 22$, these values increase to 2.22 and
2.66~mas y$^{-1}$ respectively for $22 \le I < 24$.

The VPD allows us to mark out some potentially interesting kinematic
outliers for further consideration. We have drawn a more or less
arbitrary section around the main bulk of the proper-motions for
identification in the reduced proper-motion and color-magnitude
diagrams (see below). There are, however, other interesting objects
that fall within the outliers boundary, and these will be discussed
later.

Recently, Oppenheimer et al. (\cite{oppenheimer}) found a large
population of white dwarfs with kinematics representative of the
Galactic Halo. These objects were pre-selected from a photographic
proper-motion survey using UK~Schmidt plates. They used the so-called
'reduced-proper-motion' (RPM) parameter H, used for the first time by
Luyten many years ago. The interesting property of this parameter is
that while it is directly observable once we have proper-motions and
photometry, it is mostly an indicator of luminosity for distinct
stellar populations. This parameter is defined as:

\[
H_m = m + 5 \log \mu + 5 = M + 5 \log V_{tan} - 5 \log K
\]
where m and $\mu$ are the apparent magnitude and (absolute)
proper-motion, and M and $V_{tan}$ are the absolute magnitude (in the
same band) and total tangential velocity for the star, while $K$ is a
constant.

As it can be seen from its definition, different stellar populations
(with characteristic luminosities and kinematics) will occupy a
different locus on, e.g., an H {\it vs.} color diagram, hence H can be
used to distinguish distinct populations from its position on the RPM
diagram.

\begin{figure}
\resizebox{\hsize}{!}{\includegraphics{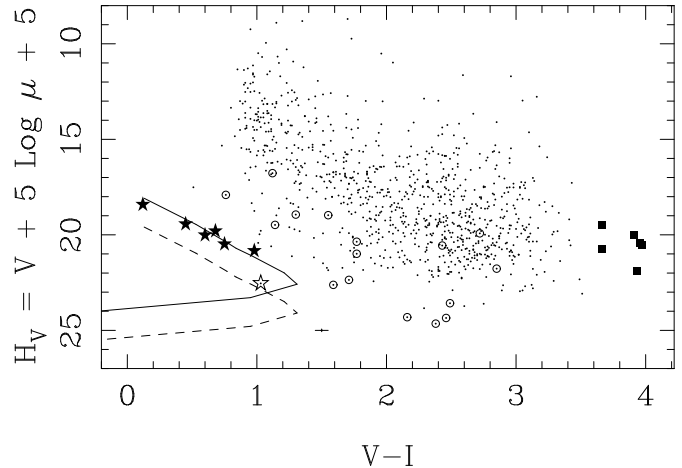}}
\caption{Reduced proper-motion diagram for the field stars toward
NGC~6397. Same symbols as for Fig.~\ref{vpd}. Solid and dotted lines
are the theoretical locus for 0.8~$M_{\sun}$ white dwarfs of different
cooling ages, from the models by Chabrier et al.~(\cite{chabrier}) for
a tangential velocity of 60 and 120~km~s$^{-1}$ respectively. The six
solid stars match well the theoretical expectation for a population
typical of the Thick-Disk, whereas there is one object that seems to
have Halo-like kinematics. Size of the errors is marked by the plus
sign.}
\label{rpmfield}
\end{figure}

Fig.~\ref{rpmfield} shows the $H_V$ {\it vs.} V-I diagram for our {\it
HST} sample. The same objects found on the VPD have been marked on the
RPM. Indeed, the RPM diagram allows us to clearly distinguish some
objects that did not stand-out in the VPD (e.g., the red stars or some
of the faint blue objects), and vice-versa (the general kinematic
outliers). The properties of the most interesting objects found (i.e.,
the extremely red stars, and the faint blue stars) are given in
Table~\ref{extremobj}.

As it can be seen in Fig.~\ref{rpmfield}, there is a group of six very
red stars with $21 < I < 24$ and colors $V-I > 3.6$. These objects are
interesting because, with their faint magnitudes and red-colors, they
could be representatives of a very faint and nearby sample of
low-luminosity late-type dwarfs or even brown-dwarfs. There is also a
group of six faint blue stars ($22 \le I < 25$, $V-I \le 1.0$) with
intermediate kinematic, and one fainter blue object with even more
extreme kinematics (labeled as 'Extreme' on
Table~\ref{extremobj}). These faint blue objects, as will be argued
below, are part of two distinct populations, belonging to the
Thick-Disk of the Galaxy and to the Halo respectively, and represent
the late-stages in the evolution of the precursors of these stars when
the Galaxy was formed. The properties of all these objects (i.e., the
extremely red stars, and the faint blue stars) are given in
Table~\ref{extremobj}.

\begin{table}
\caption{Properties of the red and faint blue stars in the field
population.}

\label{extremobj}
\begin{tabular}{@{}lccccc}
\hline
\hline
Objects &   I    &   V-I   & $\mu_l$      & $\mu_b$      \\
        &  mag   &   mag   & mas y$^{-1}$ & mas y$^{-1}$ \\
\hline
Red     & 21.02  &  3.66   &  -8.65       &  3.02        \\
        & 21.06  &  3.97   & -12.42       &  1.59        \\
        & 22.42  &  3.66   &  -8.29       &  2.26        \\
        & 22.52  &  3.91   &  -1.10       &  5.12        \\
        & 23.28  &  3.96   &  -0.92       & -4.26        \\
        & 23.63  &  3.93   &  -6.78       &  2.60        \\
\hline
Blue    & 22.79  &  0.60   & -18.24       & 10.71        \\
        & 23.28  &  0.12   &  -6.24       & -7.91        \\
        & 24.15  &  0.68   &   9.71       &  1.90        \\
        & 24.34  &  0.45   &   8.12       & -2.57        \\
        & 24.51  &  0.98   &  10.53       &  5.26        \\
        & 24.89  &  0.75   &   9.28       & -1.09        \\
\hline
Extreme & 24.03  &  1.03   & -20.07       &-24.13        \\ 
\hline
\hline  
\end{tabular}
\end{table}

An inspection of Table~\ref{extremobj} indicates that the red stars
are actually not very nearby: If they belong to the disk population
(as predicted from all Galactic models, see e.g. Castellani et al.\
(2001)), then their kinematics (assuming a tangential velocity of
40~km~s$^{-1}$, typical of this population) indicates distances larger
than 600~pc, and absolute magnitudes brighter than $M_I \sim 12.0$,
characteristic of M5 to M6-type dwarfs (Gizis~1997; Leggett~1992, her
Fig.~1). Their colors also indicate similar distances and
luminosities. For example, adopting the color-absolute magnitude
relationship given by Gizis (\cite{chabrier}, his Table~4 and Fig.~11)
or Leggett (\cite{legget92}, her Fig.~15), we have for $V-I \sim 4.0$
a typical $M_V \sim 16.5$ or $M_I \sim 12.5$. We conclude that none of
these red stars are very nearby nor particularly faint. On the other
hand these objects can not be extragalactic, because they exhibit
motions that are significantly different from zero, within our
estimated 2~mas~y$^{-1}$ error budget (see Table~\ref{proppm}).

The object with the most extreme-kinematic (the star far to the left
on the VPD, see Fig.~\ref{vpd}) has a V-I=~2.85 and I=~19.56. The
absolute magnitude for this color is predicted in the range $12.0 \le
M_V \le 13.5$, depending on metallicity (Gizis 1997). Thus, for a
total proper-motion of 74.43~mas~y$^{-1}$ one derives a tangential
velocity on excess of 200~km~s$^{-1}$, i.e. more than five times the
expected tangential velocity of a typical Disk star. This object is
thus quite interesting for follow-up spectroscopy as a potential Halo
subdwarf (the same is true for the other kinematic outliers, 16
objects in total, located outside the polygon on
Fig.~\ref{vpd}). Alternatively, if we insist on this object having a
tangential velocity of 40~km~s$^{-1}$, one derives a distance of about
120~pc, leading thus to an absolute magnitude of $M_V \sim 17.0$, {\it
i.e.}, more than three magnitudes fainter than normal Disk stars of
the same color, an unlikely solution.

The sample of faint blue objects represents an interesting population
that deserves further attention, in view of the recent discovery by
Oppenheimer et al.~(\cite{oppenheimer}) mentioned
before. Oppenheimer's study was limited to much brighter magnitudes
(with $R \le 19.8$) than our study ($20 \le I \le 25$, see
Fig.~\ref{cmd}), but they have spectral confirmation. Our objects are
much fainter, and it is unlikely that we will have any confirmatory
spectroscopy on them until the large next-generation telescopes of
30~m or more become available. Table~\ref{extremobj} shows that all
these objects have motions that differ by more than 2.5 times from
both the mean motion of the field or the cluster, even considering the
large width of the field's proper-motion dispersion of more than
4~mas~s$^{-1}$ given in Table~\ref{proppm}.

Even though we can not hope to have spectroscopy on these objects yet,
we can use the RPM diagram, along with theoretical models to, at
least, compare their position on this diagram with what we would
expect for ancient white dwarfs. On Fig.~\ref{rpmfield} we have
superimposed the latest theoretical cooling sequences for old white
dwarfs from Chabrier et al.~(\cite{chabrier}). The continuous line is
for a population with a mean tangential speed of 60~km~s$^{-1}$,
characteristic of the Thick-Disk (TD model), whereas the dashed line
is for a speed of 120~km~s$^{-1}$, typical of the Halo (Halo
model). As can be seen the brightest blue stars fall very close to the
expected location for old white dwarfs from the Thick-Disk. The
faintest point however could, in principle, be fit by either
model. However, we argue that this object is better fit by the Halo
model rather than the TD model. Indeed, if this was a Thick-Disk star,
then the corresponding (cooling) age (i.e., since it left the
main-sequence) from the theoretical sequence (almost 14~Gyr), would
imply that the Thick-Disk is quite older than what most models
predict. It will also imply an absolute magnitude for this object of
$M_V \sim 17.6$, thus becoming the faintest degenerate objects ever
found (as an example ESO~439-26, the faintest spectroscopically
confirmed WD known has $M_V \sim 17.4$, Ruiz et al.~1995). Instead, if
this object is placed on the Halo model, then its derived cooling age
is 11~Gyr, fully consistent with the age for the Galactic Halo, and
with a reasonable luminosity of $M_V \sim 15.8$. We must emphasize
here that the derived tangential velocities for these objects are used
only as an indication of the population to which they belong, but
these velocities are not used to determine distances directly (see
Section~4).

\subsection{Comparison to local and simulated samples}

\begin{figure}
\resizebox{\hsize}{!}{\includegraphics{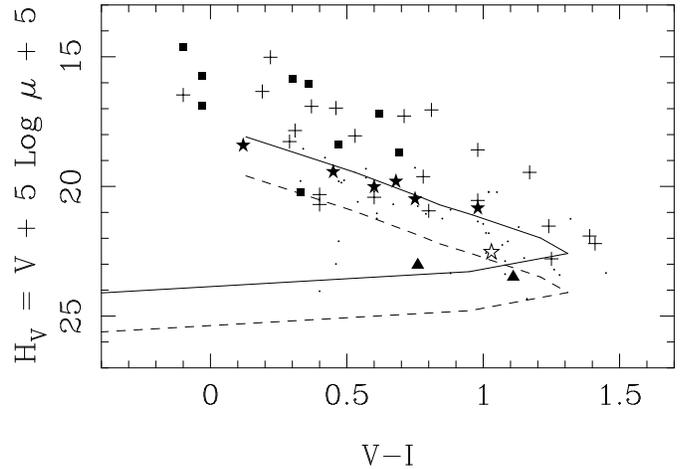}}
\caption{Reduced proper-motion diagram for local samples of white
dwarfs selected according to different criteria. Small dots are from
the Legget, Ruiz and Bergeron~(\cite{legget98}) sample with $\mu >
0.8$~arc-sec~y$^{-1}$. Plus and solid-square symbols are for the
sample from Ruiz and Bergeron~(\cite{ruiz01}) with $\mu >
0.2$~arc-sec~y$^{-1}$ and $\mu \le 0.2$~arc-sec~y$^{-1}$
respectively. Open and solid stars is our sample (same as in
Fig.~\ref{rpmfield}), and triangles are the low-luminosty white dwarfs
ESO~439-26 and CE~51 with extreme kinematics, as described in the
text. Solid and dashed lines are the same as in Fig.~\ref{rpmfield}.}
\label{rpmlocal}
\end{figure}

Before we can proceed with any certainty on the probable
identification of the faint blue objects as belonging to the stellar
Thick-Disk or Halo components of our Galaxy as described in the
previous paragraph, we need to compare this sample of putative white
dwarfs with the kinematics of local samples of (Disk) white dwarfs. A
sample with well defined selection criteria, which has been used for a
determination of the (Disk) white dwarf luminosty function is provided
by Legget, Ruiz, and Bergeron (\cite{legget98}, LRB98
thereafter). Fig.~\ref{rpmlocal} shows the locus in the RPM diagram of
the LRB stars in comparison with our sample. At first it would seem as
if there is no difference between the two samples, thus implying that
we are actually looking at normal Disk stars. However, there is a big
difference in the two samples: The LRB sample has a strong kinematic
bias in favor of stars with quite large proper-motion ($\mu >
0.8$~arc-sec~y$^{-1}$), and thus the mean tangential velocity of the
LRB89 sample is not representative of the Disk population of white
dwarfs as a whole, but rather it overestimates it. This then displaces
the mean locus of these objects on the RPM to larger values of $H_V$,
making them appear similar to that of our sample (we must note that
our sample has not been pre-selected according to a large
proper-motion). The mean tangential velocity for the LRB8 sample is
113~km~s$^{-1}$ with a dispersion of 73~km~s$^{-1}$. It would of
course be interesting to compare directly our sample with that of Disk
white dwarfs that have not been selected according to kinematics. Such
a sample does exist (Holberg, Oswalt, and Sion 2001, HOS01
thereafter), but unfortunately there are are no V-I colors available
for these stars which would allow us to make a direct comparison with
our objects. From HOS01's (volume-limited) complete sample of white
dwarfs closer than 13~pc one finds that their mean tangential velocity
is only $45 \pm 5$~(m.e.)~km~s$^{-1}$ (this is actually based on 43
out of their 49 stars closer than 13~pc due to the lack of
proper-motions for a few of them). We note here that volume-limited
samples of main-sequence stars from Hipparcos give values close to
40~km~s$^{-1}$ (Dehnen and Binney 1998), i.e., similar to those found
for Holberg's sample (actually one would expect white dwarfs to have
larger motions since they are older than the disk, and much of the
velocity scattering is generated in the first Gyr (Wielen
1977). Indeed, numerical simulations using Reyl\'e's model (private
communication, model described in Reyl\'e, Robin, and Cr\'ez\'e 2001),
for the same Galactic location as our field, and with the same
magnitude, color, and error constraints as those of our sample give a
$<V_{tan}> = 45.8 \pm 1.9$~(m.e.)~km~s$^{-1}$, in excellent agreement
with the results from HOS01 sample of nearby Disk white dwarfs. From
this analysis we see that the mean tangential velocity for our
possible Thick-Disk white dwarfs is more than $7 \sigma$ away from the
mean of the local (simulated) sample. A further indication of the
effect of kinematic selection is provided by the sample of Ruiz and
Bergeron (\cite{ruiz01}, RB01 thereafter). They have published a
sample of new nearby spectroscopically confirmed white dwarfs,
selected from their own proper-motion survey. For their sample with
$\mu > 0.2$~arc-sec~y$^{-1}$ one finds $<V_{tan}> = 60 \pm
8$~(m.e.)~km~s$^{-1}$, while for $\mu \le 0.2$~arc-sec~y$^{-1}$ one
finds $<V_{tan}> = 43 \pm 12$~(m.e.)~km~s$^{-1}$ (this last sample is
incomplete in proper-motion, so this value is probably an upper
limit).

While comparing samples of putative disk white dwarfs to those
presented in this paper, one also has to keep in mind that it is not
unlikely that a few of the 'local' samples may also have
representatives from the Thick-Disk and Halo
(Fig~\ref{rpmlocal}.). This is the case, for example of ESO439-26
(with $M_I = 16.29$ and $V_{tan} = 78$~km~s$^{-1}$, Ruiz et al.~1995)
or CE51 (with $M_I = 16.47$ and $V_{tan} = 68$~km~s$^{-1}$, RB01). The
mean tangential estimates derived from these local samples will be
biased toward higher values due to these kinematic
interlopers. Unfortunately, spectroscopy does not help here because
all white dwarfs, irrespective of the metallicity of the parent
population, look alike.

\begin{figure}
\resizebox{\hsize}{!}{\includegraphics{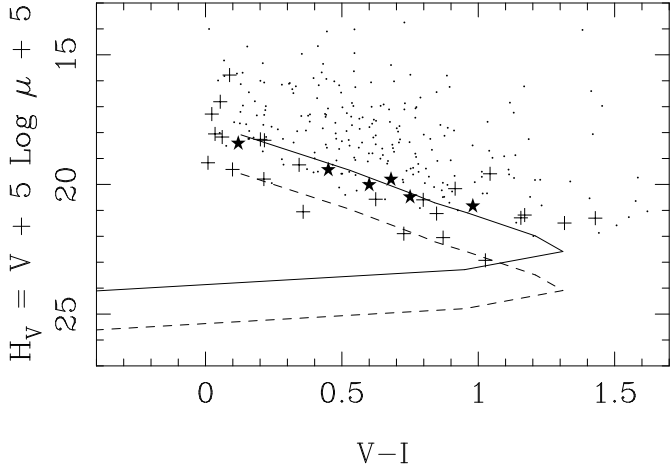}}
\caption{Reduced proper-motion diagram for simulated samples of white
dwarfs selected according to the very same criteria as those of our
sample. The simulatiosn are based on the Besan\c{c}on Galactic model
of stellar population synthesis (Celine Reyl/'e, private
communication). Small dots are (simulated) Disk white dwarfs, plus
symbols are (simulated) Thick-Disk stars, and star symbols are the
objects found in this study. As it can been seen, our objects trace
well the predicted locus for Thick-Disk stars as inferred from the
model.}
\label{rpmreyle}
\end{figure}

We can finally compare the locus of our sample on the RPM diagram to
that predicted by Galactic models. Here we use the predictions from
the model by Reyl\'e et al.~(\cite{reyle}, and private communication)
based on the Besan\c{c}on model of stellar populations in the Milky
Way, which has been (succesfully) used to identify the kinematics of
Oppenheimer's sample with that of Thick-Disk stars (see also Reid et
al. 2001). Model simulations where run for the same Galactic location
as that of our sample, with the same magnitude, color and
proper-motion limits, as well as the same observational errors in all
these quantities. The field-of-view of the simulation was larger than
that of our sample in order to have enough (simulated) stars in the
comparison. Fig.~\ref{rpmreyle} shows the locus for our stars, and
those predicted by the model. As it is obvious from this figure, the
location of our stars is totally in agreement with the predictions
from the model for Thick-Disk stars. Disk stars on the other hand
exhibit values with smaller tangential velocity (smaller $H_V$), as it
has been discussed before. Of course all this evidence is, in a way,
indirect (even if, as dicussed before, we had spectroscopy), and one
could still insist that these stars are actually members of the Disk
population. But, in this case, how can one explain that they obey such
a distint kinematic behavior in the RPM diagram in comparison with the
normal Disk stars?  This obviously points to these objects as beeing
part of a separate higher-velocity component. We thus conclude that,
based on the evidence we have, these few blue faint stars are very
likely true representatives of the Thick-Disk of our Galaxy. The
implications for this will be discussed in the following Section.

\begin{figure}
\resizebox{\hsize}{!}{\includegraphics{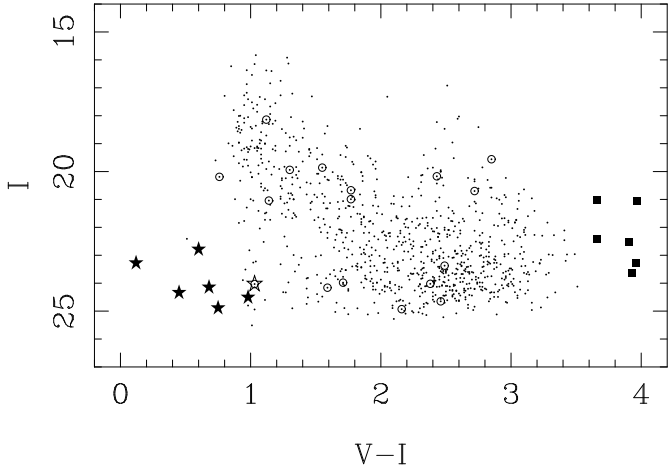}}
\caption{Color-magnitude diagram for the whole sample of field stars,
adopting the same symbols as for Figs.~1 and~2. The separation between
distinct kinematic groups becomes fuzzy.}
\label{cmd}
\end{figure}

A final point is important when comparing the results from this paper
with those found by MM00 for the HDF-S\&N. Fig.~\ref{cmd} shows the
calibrated color-magnitude diagram for our sample. As it can be seen,
the faint blue objects are much less well defined, becoming mixed with
the rest of the distribution, and making it difficult to distinguish
between different kinematic components, a point to which we will come
back in the next Section. The outliers on the VPD diagram (open
circles) also become well mixed with the rest of the distribution, and
only the red objects stand-out by virtue of their color-selection
criteria.

\section{Discussion}

It is interesting to 'weight' (literally) the consequences of the
presence of two kinematic components on the faint blue population
described above. First of all, is the discovery of one possible truly
Halo white dwarf (the open star on Figs.~1 and~2) consistent with
previous findings? As it was shown by MM00, the total mass contributed
by a Halo-like population is given by:

\begin{eqnarray}
M(r_{max}) & \cor & <M_{WD}> \times N_{WD}  =  \nonumber \\
& & 8.46 \times 10^{-8} \Omega \times \rho_{\sun} \int_{0}^{r_{max}}
\left( \frac{R_{\sun}}{R} \right)^3 r^2 \, dr
\label{eqn4}
\end{eqnarray}
where $\rho_{\sun}$ is the local mass density of this component,
$\Omega$ is the field-of-view in arcmin$^2$, R is the galactocentric
distance (in pc), $R_{\sun}$ is the Solar Galactocentric distance
(assumed to be 8,500~pc), r is the heliocentric distance (pc), and
$M(r_{max})$ is the mass (in $M_{\sun}$) contributed by this component
out to a Heliocentric distance $r_{max}$ (also in pc), while $N_{WD}$
is the number of observed white dwarfs of mean mass $<M_{WD}>$. If we
compute the ratio of $M(d_{max})$ between, e.g., the Hubble Deep
fields South and this field close to NGC~6397, then we find that, {\it
to the same depth in apparent magnitude}, the expected number of Halo
white dwarfs would scale as $N_{WD}(N6397) \sim 1.11 \times
N_{WD}(HDF-S)$, taken into account the different fields-of-view of the
two samples. Now, down to $I \sim 25$ we see from Fig.~3 on MM00 that
we have two possible Halo white dwarfs on HDF-S, whereas we have found
one based on its kinematic properties on the field of NGC~6397. No
doubt, it is small numbers, but at least the two studies (MM00 based
only on differential star counts, the present one on kinematics) are
consistent with each other. Note also that the HDF images go about 5
magnitudes fainter than the ones on the present study. We conclude
thus that there is consistency between these results.

In our study we seem to find a population of stars somewhat brighter
than those detected by MM00, with kinematics indicative of the
Thick-Disk. However, there is no indication for these objects on the
HDF-S\&N sample by MM00, which seems to be inconsistent with the
findings of the present paper. Here we should point out that there was
no kinematic information for the sample used by MM00 and, thus, it was
not possible to kinematically distinguish Thick-Disk from Halo-like
objects. Therefore, it is in principle possible that {\it some} of the
blue objects on MM00 could actually be Thick-Disk stars. However, it
turns out that the expected density of these latter objects is much
smaller in the HDF-S\&N because of their spatial location in the
Galaxy. If we adopt a density law characteristic of the Thick-Disk of
the form:

\begin{equation}
\rho(\vec r)  = \rho_{\sun}  \exp^{-|z-z_{\sun}|/h_z} 
\exp^{-(R_{pl}-R_{\sun})/h_R}
\label{eqn5}
\end{equation}
where $\rho(\vec r)$ is the stellar densities at Galactic position
$\vec r$ whereas $\rho_{\sun}$ is the density in the solar
neighborhood, $h_z$ and $h_R$ are the scale height and length of this
component respectively, z is the distance from the Galactic plane,
$z_{\sun} \sim 7$(pc) is the Sun displacement with respect to the
Galactic plane, and $R_{pl}$ is the distance from the Galactic center,
as measured on the plane of the Galaxy, we find, using the analogous
of equation~\ref{eqn4} for this density law, that the ratio between
the expected number of objects on the HDF-S\&N and the field under
consideration is only 0.29 and 0.16 for HDF-S and HDF-N
respectively. So, if we have 6 Thick-Disk stars on the N6397 field, we
should only expect to see 1 or 2 such objects on HDF-S and 1 or none
on HDF-N, for a reasonable choice of $h_z$ (760~pc) and $h_R$
(2500~pc) in the above equation. It is interesting to notice that,
above $V \sim 25$ (i.e. the same magnitude range as our stars, see
Fig.~\ref{cmd}), there is indeed only one blue object on HDF-N and two
blue objects on the HDF-S (see Fig.~3 on MM00). These objects have
magnitudes and colors similar to those of our kinematically selected
Thick-Disk sample in NGC~6397, and in agreement with the above
numbers. However, lacking kinematics for the HDF sample, we can not
advance much more in this analysis (see below).

\subsection{Density estimates}

We can use the six stars identified as members of the Thick-Disk to
compute the stellar and mass density associated to this component. For
this purpose we use the generalized $1/V_{\rm max}$ method proposed by
Tinney et al.~(\cite{tinney}). This method extends the classical
$1/V_{\rm max}$ to the case of non-(spatially) uniform samples, such
as those found here. The method defines a (generalized)
density-weighted volume $V_{\rm gen}$ as:

\begin{equation}
V_{\rm gen}  = \int_{Vol} \frac {\rho(\vec r)}{\rho_{\sun}} \, dVol
\label{eqn6}
\end{equation}
where $\rho(\vec r)$ and $\rho_{\sun}$ are the stellar densities at
position $\vec r$ and in the solar neighborhood respectively, and
the integral is carried out over the volume available to the sample,
given the observational constraints.

Then, the total stellar density is given, just as in the classical
method, by:

\[
\rho_{\sun} = \sum_{i} \frac{f_i}{V_{{\rm gen}_i}}
\]
where the factors $f_i$ account for the incompleteness of the sample
($f_i \ge 1$), and the sum is carried over the individual data
points. Meanwhile, the standard deviation on the derived overall
density is given by:

\[
\sigma_{\rho_{\sun}} = \sqrt { \sum_{i} \frac{f_i}{V_{{\rm gen}_i}^2} }
\]
The volume over which the integral in equation~\ref{eqn6} is to be
extended is determined by the minimum and maximum distances ($r_{\rm
min}$ and $r_{\rm max}$ respectively) over which any star can
contribute to the sample (see, e.g., M\'endez and Ruiz~2001).

If we have a sample with a lower proper-motion limit $\mu_l$ and a
faint apparent magnitude $m_f$, $r_{\rm max}$ is given by:

\[
r_{\rm max} = min [p^{-1} (\mu/\mu_l) ; p^{-1} 10^{0.2(m_f-m)}]
\]
where $p$ is the parallax, $\mu$ is the proper-motion and m the
apparent magnitude.

Similarly, if the sample is only complete to an upper proper-motion
limit $\mu_u$ and a bright apparent magnitude $m_b$, the minimum
distance for inclusion would be:

\[
r_{\rm min} = max[p^{-1} (\mu/\mu_u) ; p^{-1} 10^{0.2(m_b-m)}]
\]
For our sample, the value for $\mu_l$ is determined by the
observational errors. As mentioned in Sect.~2.2, our observational
errors are on the order of 2~mas~y$^{-1}$. Therefore, one could
consider to have a firm detection for, say, three times the error, or
6~mas~y$^{-1}$, which we will adopt as our value for $\mu_l$. Objects
with motions larger than that would qualify within our selection
criteria. Similarly, the fastest-moving object in the sample (see
Fig~\ref{vpd}.) has a total proper-motion of nearly 75~mas~y$^{-1}$,
value thus adopted for $\mu_u$. It is very likely that the detection
limit is larger than this, and it is determined by the epoch
difference used to obtain the proper motions and the confusion
limit. However, we have made tests and found that, even if one where
to adopt $r_{\rm min} = 0$, our results will change by less than
0.1\%. The apparent magnitude limits were chosen as $m_b = 15$ (see
Fig.~\ref{cmd}), and $m_f = 25$ (see Table~1 on KACP98).

For each of the six putative Thick-Disk stars identified in
Table~\ref{extremobj}, we computed their absolute magnitudes in the I
pass-band from their observed V-I color, by interpolation on Table~3
of Chabrier et al.~(\cite{chabrier}). The I and $M_I$ were then used
to obtain photometric parallaxes. We must emphasize here that the
kinematics of the sample is not used at all to determine photometric
distances or absolute magnitudes, rather the kinematics is used as an
indication of belonging to a given population, which then allows us to
adopt the correspondig density laws (e.g., equation~\ref{eqn5}) for
calculating the effective sampling volume
(equation~\ref{eqn6}). Therefore our results are not dependant on the
exact value adopted for $V_{tan}$ on either the Thick-Disk or the
Halo.

With the above values, the limiting distances $r_{\rm min}$ and
$r_{\rm max}$ were computed, and the integral on equation~\ref{eqn6}
was evaluated by direct integration, using the adopted Thick-Disk
density law, equation~\ref{eqn5}. In this case, equation~\ref{eqn6}
can be written as:

\begin{eqnarray}
V_{\rm gen} & = & 8.46 \times 10^{-8} \Omega \int_{r_{min}}^{r_{max}}
\exp^{-|z-z_{\sun}|/h_z} \, \exp^{-(R_{pl}-R_{\sun})/h_R} r^2 \, dr 
\nonumber \\
& = & V_{\rm max} \times \Re
\label{eqn7}
\end{eqnarray}

Where $V_{\rm max}$ is the classical maximum allowed volume for a
magnitude and proper-motion selected sample (see, e.g., M\'endez and
Ruiz 2001), and $\Re$ is a (dimensionless) correction factor which
accounts for the non-homogeneous distribution of stars in space (we
shall see that, in any case, this correction factor is relatively
small, close to one, for our sample). Both terms are given,
respectively, by:

\begin{equation}
V_{\rm max} = 8.46 \times 10^{-8} \times \Omega \times
\left( \frac{r_{max}^3-r_{min}^3}{3} \right)
\label{eqn7_1}
\end{equation}
and

\begin{equation}
\Re = \frac{\int_{r_{min}}^{r_{max}} \exp^{-|z-z_{\sun}|/h_z} \,
\exp^{-(R_{pl}-R_{\sun})/h_R} r^2 \,
dr}{\left( \frac{r_{max}^3-r_{min}^3}{3} \right) }
\label{eqn7_2}
\end{equation}

The density law in equation~\ref{eqn7} or equation~\ref{eqn7_2} has
two parameters, namely $h_z$ and $h_R$, which introduce some ambiguity
in the computation of the integral. We have thus computed the integral
(and the corresponding correction factor R) for several combinations
of parameters in the range $760 < h_z < 1300 \, {\rm pc}$ and $2500 <
h_R < 3500 \,{\rm pc}$, and have chosen the extreme values as
representative of the 'modeling errors' involved in the computation.
Finally, the factors $f_i$ have been computed as the reciprocal of the
completeness fraction as inferred from the artificial-star experiments
reported by KACP98 (their Table~1). Table~\ref{rminmax} summarizes all
the adopted and derived values for the six Thick-Disk and one Halo
stars from Table~\ref{extremobj}. The cooling age on the second column
of Table~\ref{rminmax} is the age (since becoming a white dwarf)
inferred from the same table on Chabrier et al.~(\cite{chabrier}) used
for determining the absolute magnitudes.

Because of the Galactic location of this field, it is easy to show
that, with small errors, the ($\mu_l,\mu_b$) proper-motions are
projected almost directly into the V (along Galactic rotation) and W
(towards the North Galactic Pole) velocity components
respectively. These velocities, computed using the distance given in
the fourth column of Table~\ref{rminmax}, are indicated, for
reference, in the seventh and eigth columns of the same
Table. Finally, in the last column we give the approximate tangential
velocity (neglecting the small contribution from the U component of
the velocity towards the Galactic center). As it can be seen from the
Table, the mean tangential velocity is close to $65~kms^{-1}$ for the
Thick-Disk stars, and more than $100~kms^{-1}$ for the one possible
Halo white dwarf, in agreement with the value found from
Fig.~\ref{rpmfield}, and larger than the value expected for normal
nearby Disk white dwarfs, as already discussed in Section~3.1.

\begin{table}
\caption{Photometric distances and derived parameters for the faint
blue stars on Table~\ref{extremobj}.}

\label{rminmax}
\begin{tabular}{@{}lcccccccc}
\hline
\hline
Objects & Age & $M_I$ & Dist & $r_{\rm min}$ & $r_{\rm max}$ & ~V       &  ~W       &  $~V_{tan}$\\
        & Gyr &  mag  &  pc  & pc            & pc            & $kms^{-1}$ &  $kms^{-1}$ &  $kms^{-1}$\\
\hline
Blue    & 3.9 & 13.62 &  682 & 192           & 1888          & -59      &  35       &  69\\
	& 1.0 & 12.44 & 1472 & 198           & 2474          & -43      & -55       &  70\\
        & 4.9 & 13.85 & 1148 & 151           & 1698          &  53      &  10       &  54\\
        & 2.6 & 13.22 & 1675 & 190           & 2270          &  64      & -20       &  67\\
        & 10.2& 14.67 &  929 & 146           & 1164          &  46      &  23       &  51\\
        & 5.8 & 14.06 & 1466 & 183           & 1542          &  64      &  -8       &  65\\
\hline
Extreme & 10.6& 14.80 &  702 & 294           & 1097          & -67      & -80       & 104\\ 
\hline
\hline  
\end{tabular}
\end{table}

Equation~\ref{eqn7_2} has thus been integrated using the parameters
specified in Table~\ref{rminmax}. The adopted incompleteness factors
and the extreme values derived for the integral in $\Re$ are given in
Table~\ref{intminmax}. $\Re_{min}$ was derived adopting $h_z =760 \,
{\rm pc}$ and $h_R = 3500 \, {\rm pc}$, whereas $\Re_{max}$ was
derived using $h_z =1300 \, {\rm pc}$ and $h_R = 2500 \, {\rm pc}$. In
the case of the Halo star, instead of using a double-exponential
density law, we have adopted an $R^{-3}$ law similar to that used in
MM00, which leads to equation~\ref{eqn4} above (to be more precise
$V_{\rm gen}$ is given by equation~\ref{eqn4} divided by the local
stellar density $\rho_{\sun}$, while the definition of $\Re$ is
similar to that adopted for a double exponential. In this case
however, the free parameter in the density law is the axial ratio of
the (flattened) Halo, which was varied in the interval 0.8 (oblate
spheroid, $\Re_{min}$) to~1.0 (round spheroid, $\Re_{max}$)).

\begin{table}
\caption{Adopted incompleteness fraction and derived values for
$\Re$.}

\label{intminmax}
\begin{tabular}{@{}lccc}
\hline
\hline
Objects & f & $\Re_{min}$  & $\Re_{max}$\\
\hline
Blue    & 1.085 & 0.9813 & 1.3307 \\
	& 1.103 & 0.9691 & 1.4440 \\
        & 1.166 & 0.9854 & 1.2957 \\
        & 1.212 & 0.9741 & 1.4048 \\
        & 1.290 & 0.9947 & 1.1989 \\
        & 2.167 & 0.9868 & 1.2654 \\
\hline
Extreme & 1.153 & 1.3206 & 1.3212 \\
\hline
\hline  
\end{tabular}
\end{table}

If ($\rho_{\sun_{min}}$, $\sigma_{\rho_{\sun_{min}}}$) and
($\rho_{\sun_{max}}$, $\sigma_{\rho_{\sun_{max}}}$) are the densities
and its error derived from $R_{min}$ and $R_{max}$ respectively,
then the final value for the total density and its error was
calculated from:

\[
\rho_{\sun} = \frac{\rho_{\sun_{min}}/\sigma_{\rho_{\sun_{min}}}^2 +
\rho_{\sun_{max}}/\sigma_{\rho_{\sun_{max}}}^2}
{1/\sigma_{\rho_{\sun_{min}}}^2 + 1/\sigma_{\rho_{\sun_{max}}}^2}
\]
,and,

\[
\sigma_{\rho_{\sun}} = \frac{1}
{\sqrt {1/\sigma_{\rho_{\sun_{min}}}^2 + 1/\sigma_{\rho_{\sun_{max}}}^2}  }
\]
With these equations, and the values given in Table~\ref{intminmax}
we obtain, for the Thick-Disk white dwarf stars a value of:

\begin{eqnarray}
\rho_{{\sun}_{WDTD}} & = &(13.82 \pm 4.39) \times 10^{-3} \, stars/pc^3 
\nonumber \\
& = & (11.06 \pm 3.51) \times 10^{-3} \, M_{\sun}/pc^3 \nonumber
\end{eqnarray}
For the one star that probably belongs to the Halo, the same
computation above yields:

\begin{eqnarray}
\rho_{{\sun}_{WDH}} & = & (5.31 \pm 3.50) \times 10^{-3} \, stars/pc^3
\nonumber \\
& = & (4.25 \pm 2.80) \times 10^{-3} \, M_{\sun}/pc^3 \nonumber
\end{eqnarray}
Obviously, we have much larger errors in this latter case due to the
use of only one object. In both cases we have assumed a mean mass per
white dwarf of 0.8~$M_{\sun}$.

Because the correction factors $\Re$ given in
Table~\ref{intminmax} are all close to one, one might as well have
computed the stellar densities using the classical $1/V_{\rm max}$
method, without correcting for the variation of stellar density as a
function of distance away from the Sun. In this case, one does not
need to adopt any density law whatsoever, and the method is completely
model independent. It is thus interesting to compare the previous
values with the ones that one would have derived in this
case. Dropping the correction factor $\Re$ in equation~\ref{eqn7}, one
finds:

\begin{eqnarray}
\rho_{{\sun}_{WDTD}} & = &(13.82 \pm 4.39) \times 10^{-3} \, stars/pc^3 
\nonumber \\
& = & (15.66 \pm 3.51) \times 10^{-3} \, M_{\sun}/pc^3 \nonumber
\end{eqnarray}
And for the Halo star:

\begin{eqnarray}
\rho_{{\sun}_{WDH}} & = & (7.02 \pm 6.54) \times 10^{-3} \, stars/pc^3
\nonumber \\
& = & (5.62 \pm 5.23) \times 10^{-3} \, M_{\sun}/pc^3 \nonumber
\end{eqnarray}

As we see, the densities are slightly overestimated, but also the
errors are larger. Still, we obtain basically the same results using
either method. This is important, because it implies that, regardless
of the details of the (model) density law adopted, the stellar and
mass densities are indeed quite large. The implications of these
results are discussed in the next Section.


\section{Conclusions}

The stellar and mass densities derived above imply that the Thick-Disk
white dwarf stars, a population {\it unaccounted for in the Hubble
Deep Fields South and North due to their high Galactic latitude},
contributes to about 85\% of the missing mass in the solar
neighborhood, whose density in the solar neighborhood is estimated to
be $\rho_{{\sun}_{DM}} \sim 1.26 \times 10^{-2} \, M_{\sun}/pc^3$.

Additionally, the value derived for only one possibly Halo star,
albeit having a large error, is similar to the value derived by MM00
from the larger sample of Halo white dwarfs found in the
HDF-S\&N. They found a density of $\rho_{{\sun}_{WDH}} = 4.64 \times
10^{-3} \, M_{\sun}/pc^3$. Their value, however, has to be taken with
caution because the lack of kinematic information prevented a clear
separation of the populations (compare Figs.~\ref{rpmfield}
and~\ref{cmd} above). It has been mentioned already that one might
expect one or two Thick-Disk stars on the HDF-S and one or none in the
HDF-N. If these objects were inadvertently identified by MM00 as Halo
stars, then the estimated density for these objects on the Halo would
have been too large. It is interesting to point out that on HDF-S
there are two stars on the WD tracks that are quite a bit brighter
than the rest of the blue-faint group. In the HDF-N there seems to be
a grouping of faint WDs, and a brighter one, which could also be
ascribed to the Thick-Disk (see Fig.~3 on MM00). In any case,
proper-motions for both the HDF-S\&N sample would help settle this
uncertainty.

If we combine the present results for Thick-Disk stars, with the
earlier results from MM00 for Halo stars, the overall density on WD
stars for both populations seems to be able to account for the whole
of the (local) missing mass in the Galaxy. The existence of a massive
component of Thick-Disk evolved white dwarfs has been proposed
recently by Gates and Gyuk~(\cite{gates}) to help explain some of the
(many) difficulties with a massive Halo of ancient stars. However, our
results suggest that, in addition to this Thick-Disk component, there
would also be a quite massive Halo of remnant stars. Even though their
local densities are similar, the mass locked in each population is
quite different, in account of their different stellar density
functions (double exponential {\it vs.} power-law). The simple
integration of these density functions over the whole Galaxy lead to
masses of:

\[
M_{WD-TD} = 4 \pi \rho_{{\sun}_{WDTD}} h_z h_R^3 \sim 10^{9} \,
M_{\sun}
\]
,and,

\[
M_{WD-H} = 4 \pi \rho_{{\sun}_{WDH}} R_{\sun}^3 \, \ln (R_{max}/R_{min})
\sim 9 \times 10^{10} \, M_{\sun}
\]
for the Thick-Disk and Halo components respectively. $R_{max} \sim
20$~kpc and $R_{min} \sim 1$~kpc in the last equation are the overall
extension of the Halo and the Halo core-radius respectively.

The total mass in these components is consistent with the mass
required to produce the MACHO events, as indicated by Gates and
Gyuk~(\cite{gates}), and is also not inconsistent with the total mass
estimates for the Galaxy out to a distance of 20~kpc. We also note
that with the small scale height of the Thick-Disk population ($h_z
\leq 1.5$~kpc), their contribution to the optical depth of
microlensing events would be quite small, and thus the primary sources
of the MACHO events would still be primordially the Halo white dwarfs.

The possibility of a white dwarf dominated (dark) halo has been
criticized from many fronts (see e.g. Richer~1999 and Gates and
Gyuk~2001). Among the most important criticisms is that the precursor
of these stars would have produced metals at a rate greater than
observed (Gibson and Mould~1997) or that Galaxy halos at high redshift
would be brighter than observed due to the the white dwarf precursors
(Charlot and Silk~1995). While these are no doubt important issues,
Chabrier (1999) and more recently Fontaine et al. (2001, see also
Fontaine's C.S. Beals Lecture at
http://www.astro.ubc.ca/WD\_workshop/talks/index.html) have shown that
many of these points can either be overcome, or are doubtful
criticisms of this scenario. One example is the chemical evolution
problem: The yield of a zero-metallicity stars is largely uknown, nor
has there been much modelling of the structure of these stars (e.g., a
zero metallicity star never undergoes a helium flash, which must have
an effect on the the yield of metals in the PNe phase).

Even if the problems described above with this scenario persist, as
pointed out by Lynden-Bell and Tout~(\cite{lynden}) in the first
'Russell lecture' of the new millennium, they could be avoided if the
objects found are 'pristine' white dwarfs, objects that never
underwent nuclear reactions nor followed the 'normal' path of stellar
evolution. These objects (with masses in the range $ 0.2 \leq
M/M_{\sun} \leq 1.1$) could have formed gradually and very slowly by
accretion onto planetary-sized precursor bodies in low-density regions
such as cooling flows and galaxy halos. By radiating their energy
before collapsing, these bodies would grow in mass resting on the
zero-point energy of confined electrons following the uncertainty
principle, and never getting a temperature high enough to start
nuclear burning. The difficulty with this scenario is that, although
theoretically possible, preliminary results suggest that the required
accretion rates are too slow to allow the formation of these objects
in less than a Hubble time, which keeps the puzzle regarding the
origin of these objects still open.

As we have seen before, the assignement of the bulk of the faint
blue stars found here to the Thick-Disk is actually irrelevant in the
sense that one could as well have used the classical $1/V_{\rm max}$
method to derive space densities, without regard to the origin or
association to a given stellar population of these stars. This is an
important issue in the context of whether some of the stars found by
Oppenheimer et al.~(\cite{oppenheimer}) are either Thick-Disk or Halo
stars. The controversy in this regard seems to come from using white
dwarf cooling ages while ignoring the main sequence lifetimes of the
progenitors, which for many of these stars will be
substantial. Indeed, Oppenheimer's sample seems to come from a
population of stars formed in a single burst of stars formation
between 11 and 14~Gyr ago, as indicated by the luminosity function of
these stars, which is a much more direct indication of the age of the
population than the cooling ages (see discussion on
http://research.amnh.org/users/bro). Another indication that perhaps
most of Oppenheimer's stars actually do belong to the Halo is provided
by the recent work of Koopmans and Blanford (2001) that have used a
maximum-likelihood analysis to show that these stars are consistent
with a kinematically distinct flattend Halo population at the more
than 99\% confidence level.

Albeit our objects are faint, it would be quite interesting to acquire
U-band photometry with HST. In this case, true white dwarfs will stand
out from other objects in a UBV diagram. Additionally, at least the
two brightest blue objects on Fig.~\ref{cmd} are whithin the
capabilities of current 8-m class telescopes for an spectroscopic
follow-up. This is an interesting possibility that deserves further
work.


\begin{acknowledgements}
I thank Ivan King, Jay Anderson and Adrienne Cool for making their
data available, and for helping with several questions regarding the
nature of the {\it HST} observations on NGC~6397. Linda
Schmidtobreick, Fernando Comeron, Ivan King, Dante Minniti, Fernando
Selman and William F. van Altena have provided valuable input to
improve the first draft - I appreciatee their help. I want to also
thank Celine Reyl\'e for producing the numerical simulations described
in the text using the Besan\c{c}on model of stellar population
synthesis (www.obs-besancon.fr/fesg/modele\_ang.html). Finally, the
referee Dr. B.R. Oppenheimer suggested a number of modifications to
the original manuscript that have greatly improved the final version,
I am most grateful to him for reading the paper with an open yet
rigorous mind.
\end{acknowledgements}


\end{document}